%% file: main.tex
\documentclass[aps,prev,twocolumn, preprintnumbers,floatfix,nofootinbib]{revtex4-1}
\input{Diagrams}

\usepackage{graphicx}
\usepackage{ulem}
\usepackage{amsmath, amsthm, amssymb}
\usepackage{amsfonts}
\usepackage{subfig}
\usepackage{xspace}
\usepackage{paralist}
\usepackage{multirow}
\usepackage{paralist}
\usepackage{graphicx}
\usepackage{bm}
\usepackage{times}
\usepackage{slashed}
\usepackage{color}
\usepackage{epsfig}\usepackage{dcolumn}
\usepackage{color}
\def\br{\begin{eqnarray}}
\def\er{\end{eqnarray}}
\def\be{\begin{equation}}
\def\ee{\end{equation}}

\begin{document}

\title{Search for sub-GeV Scalars in $e^+e^-$ collisions}

\author{D. Cogollo$^1$}
\author{Y.M. Oviedo-Torres $^{2,3,4}$}
\email{Corresponding Author: mauricio.nitti@gmail.com}
\author{Farinaldo S. Queiroz$^{2,3,5,6}$}
\author{Yoxara Villamizar$^{7,8}$}
\author{J. Zamora-Saa$^{2,4}$}

\affiliation{$^1$Departamento de F\'{\i}sica, Universidade Federal de Campina Grande, Campina Grande, PB, Brazil \\
$^2$Millennium Institute for Subatomic Physics at High-Energy Frontier (SAPHIR), Fernandez Concha 700, Santiago, Chile.\\
$^3$International Institute of Physics, Universidade Federal do Rio Grande do Norte, Campus Universit\'ario, Lagoa Nova, Natal-RN 59078-970, Brazil \\
$^4$Center for Theoretical and Experimental Particle Physics - CTEPP,
Facultad de Ciencias Exactas, Universidad Andres Bello, Fernandez Concha 700, Santiago, Chile\\
$^5$Departamento de Física Teórica e Experimental, Universidade Federal ´
do Rio Grande do Norte, 59078-970, Natal, Rio Grande do Norte, Brazil\\
$^6$Departamento de Fisica, Facultad de Ciencias, Universidad de La Serena,
Avenida Cisternas 1200, La Serena, Chile\\
$^{7}$ Centro de Ci\^encias Naturais e Humanas, Universidade Federal do ABC, Santo Andr\'e, 09210-580 SP, Brazil\\
$^{8}$Laboratoire de Physique Theorique et Hautes Energies (LPTHE), UMR 7589,
Sorbonne Universite et CNRS, 4 place Jussieu, 75252 Paris Cedex 05, France
}

\begin{abstract}
Light scalars that couple to leptons are common figures in beyond the Standard Model endeavors. Considering a scalar that has universal and couplings to leptons only, we compute this leptophilic scalar contribution to the $e^{-}e^{+} \rightarrow \tau^{+}\tau^{-} S $ production cross section with $S \rightarrow e^{+}e^{-}(\mu^{+}\mu^{-})$. We later compare the expected signal with recent data from the BELLE collaboration collected near the resonance $\Upsilon(4S)$ with $\mathcal{L}=626 fb^{-1}$ of integrated luminosity to place limits on the couplings-mass plane for the $4$~MeV-$6.5$~GeV mass range. We then extended this analysis to a more general one $e^{-}e^{+} \rightarrow {\ell}^{+}{\ell}^{-} S$ production cross section where $\ell=e,\mu,\tau$ with $S \rightarrow e^{+}e^{-}(\mu^{+}\mu^{-})$, showing that BELLE constitutes an excellent laboratory for light scalars, where can be derived constraints stronger than those derived stemming from the g-2 of the electron or muon.  \newline
\end{abstract}

\maketitle

\section{Introduction}
\label{introduction}

The Standard Model (SM) has successfully reproduced the precision tests carried out thus far \cite{Erler:2019hds, Beacham:2019nyx, article, Vachon:2025wub, Pasztor:2019rqu}. Understanding the origin of the masses of elementary particles has been one of the fundamental endeavors in particle physics for several decades. The Higgs mechanism, based on spontaneous symmetry breaking with the Higgs boson as a key figure, triumphed with the discovery of a new resonance with a mass of 125 GeV by both the ATLAS \cite{ATLAS:2012yve} and CMS collaborations \cite{CMS:2012qbp}. Nevertheless, it is well known that the SM does not address neutrino masses \cite{Davis:1968cp, Pontecorvo:1957cp, ParticleDataGroup:2024cfk, Mohapatra:1979ia, Mohapatra:1986bd, Acero:2022wqg} and dark matter \cite{babcock1939rotation, rubin1970rotation, rubin1978extended, zwicky1933rotverschiebung, Clowe:2006eq, Planck:2018vyg, Peebles:1982ff, XENON:2023cxc, LZ:2024zvo, Bertone:2004pz, Bertone:2016nfn, Arcadi:2024ukq}, and for many years, a hint of new physics in the muon anomalous magnetic moment persisted \cite{Muong-2:2021vma}, which motivated several new physics models.

Many of these models feature new scalar particles that are either heavier or, in some cases, much lighter than the Higgs boson \cite{Clarke:2013aya,Enberg:2016ygw,Chang:2017ynj,Liu:2018xkx,Nomura:2018yej,Wang:2018jpr,Winkler:2018qyg,Boiarska:2019vid,Chakrabarty:2019kdd,Gao:2021fyk,Hara:2021lrj,Boiarska:2019vid,Liu:2020qgx,Binder:2022pmf,Ramirez-Quezada:2022uou}. It is often assumed that the new scalar particles mix with the Higgs, and consequently the new scalar couples to SM fermions proportionally to their masses in the context of minimal flavor violation \cite{DAmbrosio:2002vsn,Cirigliano:2005ck,Kagan:2009bn}. Generally speaking, these light scalars are expected to couple to all SM fermions, but data on meson decays strongly constrain their couplings to quarks \cite{ICARUS:2024oqb}. Therefore, models with leptophilic scalars where couplings are present only to leptons, have emerged in the literature in the context of dark matter \cite{Fox:2008kb,Agrawal:2014ufa,Bandyopadhyay:2017tlq,Dutta:2017ljq,Berlin:2018bsc,Chakraborti:2020zxt,Kundu:2021cmo,Barman:2021hhg,Huang:2022ceu,Bickendorf:2022buy,John:2023ulx,Cesarotti:2024rbh}, neutrino masses \cite{Batra:2023ssq}, and the muon anomalous magnetic moment \cite{Branco:2011iw,Freitas:2014jla,Chen:2015vqy,Batell:2016ove,Liu:2020qgx,Jia:2021mwk}. This has motivated the search for light scalars in low energy accelerators featuring lepton beams which offer a relatively cleaner environment \cite{Belle:2000cnh,BaBar:2020jma,Alda:2024cxn} compared to hadron colliders \cite{Afik:2023vyl}.

In particular, these collider searches are focused on real scalar particles that couple to leptons as governed by the Lagrangian,

\begin{equation}
    \mathcal{L}_{eff}=\frac{1}{2}(\partial_{\mu} S)^2-\frac{1}{2}m_{s}^{2}S^{2}+\sum_{\ell=e,\mu,\tau} \xi_{\ell}\frac{m_{\ell}}{v}\bar{\ell}  \ell S ,
    \label{effective}
\end{equation}

where $\xi_{\ell}$ is the flavor-independent coupling to leptons ($\ell$) with mass $m_{\ell}$. In Eq.\eqref{effective}, it is assumed that the new scalar mixes with the Higgs boson, which results in its coupling to SM fermions being proportional to their masses. This scenario can arise in simple extensions of the SM and in lepton-specific Two Higgs Doublet Models \cite{Batell:2016ove}. In recent years, several groups, using different statistical methods, have derived constraints on $\xi_{\ell}$ as a function of $m_{S}$, the mass of the scalar, henceforth called ($S$) \cite{Belle:2022gbl,BaBar:2020jma}. That said, BELLE collaboration has carried out searches for light scalars in $e^{+}e^{-}$ collisions, despite being originally designed and optimized for the observation of CP violation in the B meson system \cite{Belle:2012iwr, Belle:2000cnh, Belle-II:2018jsg, Belle:2024mml, Belle:2023yoe, Belle:2024qhe}. Notably, in 2001, Belle, along with the BaBar experiment, observed CP asymmetries consistent with SM predictions, a milestone recognized with the 2008 Nobel Prize in Physics \cite{Belle:2012iwr}. As the BELLE detector features a spectrometer with a large solid angle coverage, high quality vertexing, and good charged particle tracking, and an excellent electromagnetic calorimeter, it has been able to make important contributions to charm physics, hadron spectroscopy, etc \cite{Kiesling:2013amm}.

\begin{figure}[h!]
    \centering
        \eetottP
    \caption{Dominant Feynman diagram for production of the leptophilic scalar $S$ in $e^+e^-$ collisions following Belle analysis.}
    \label{fig:diagram1}
\end{figure}

In particular, BELLE collaboration has conducted searches for such a light leptophilic scalar using collisions between 8 GeV electrons and 3.5 GeV positrons at the KEKB collider \cite{Belle:2022gbl}. The collision energy limits the maximum scalar mass that can be probed. The relevant Feynman diagram is exhibited in Fig.\eqref{fig:diagram1}. The leptophilic scalar is produced through Drell-Yan processes. Due to the Drell-Yann production and the nature of the coupling in Eq.\eqref{effective}, the dominant production channel stems from the radiation of a leptophilic scalar from a tau-lepton, primarily because of its mass. Due to kinematics, the decay channels considered are the $e^+ e^-$ and $\mu^{+} \mu^{-}$ pairs. The cross-section of $e^+e^-\to \tau^+\tau^-S$
sharply falls with the scalar mass~\cite{Batell:2016ove}. The decay channel $S \rightarrow e^+e^-$ was considered for $ 2 m_e < m_{S} \leq 2m_\mu$, while the $S$ decay into muon pairs was considered for $ 2 m_\mu < m_{S} \leq 6.5$~GeV. For $m_{S}> 2 m_\tau$, the cross-section $e^+e^-\to \tau^+\tau^-S, S \to \mu^{+} \mu^{-}$ process naturally decreases~\cite{Batell:2016ove}, but we are still able to set competitive limits on the coupling strength.

In this work, we used the latest upper limits derived by the BELLE collaboration on the cross section of the asymmetric process $e^{+}e^{-} \longrightarrow \tau^{+}\tau^{-}S$ ($S \rightarrow e^{+}e^{-}(\mu^{+}\mu^{-}))$, to establish constraints on a truly universal coupling of a hypothetical scalar $S$ with leptons. The model we investigate in this study differs subtly from the model in Eq. \eqref{effective}, as in our case, the scalar $S$ couples to leptons with an universal coupling $\lambda$, and in fact, this generates a different phenomenology. This motivates us to derive BELLE constraints in a more general setup and compare our findings with existing limits from measurements on the magnetic moment of charged leptons. 

Our paper is structured as follows: in Section \ref{g-2} we describe the status of the anomalous magnetic moment of the muon, including the most recent experimental result, which we will later use to set constraints on the plane $\lambda$-$m_{S}$, in section \ref{belledata} we describe the BELLE original analysis, in Section \ref{ouranalysis} we explain how our study differs from BELLE's and discuss our findings before concluding in Section \ref{conclusions}.

\section{The anomalous magnetic moment}
\label{g-2}

The anomalous magnetic moment (g - 2) exemplifies the predictive power of quantum field theory, as its precise measurement is vital for probing higher-order corrections in perturbation theory \cite{ParticleDataGroup:2022pth}. The g-2 muon has intrigued the community because a discrepancy between theory and data has been witnessed in the past, the hint for new physics reached $5\sigma$. The precise value of this anomaly is subject to larger uncertainties from hadronic corrections \cite{Budapest-Marseille-Wuppertal:2017okr,Ignatov:2023wma,RBC:2024fic}. In the last few years theoretical improvements have been achieved based on a technique that uses input data from other experiments \cite{Aoyama:2020ynm} and a second technique that heavily relies on computational power \cite{AOYAMA20201}, with results that are closer to the experimental measurement. Recently, the Muon g-2 collaboration have released their final measurement of the muon magnetic anomaly \cite{Muong-2:2025xyk} with improved precision. In this this report the collaboration decided to adopt a different value for the vacuum polarization contribution, and consequently, the central value for the g-2 has shifted drastically. Taking at face value their result, one can conclude that the long-standing g-2 anomaly has now disappeared. The discrepancy between the SM and experimental measurements falls within $1\sigma$. Therefore, we can conservatively use the g-2 as a probe for new physics, not as an anomaly.

Taking a step back we remind the reader that the Dirac equation predicts a muon magnetic moment, $\vec{M}=g_\mu \frac{e}{m_\mu}\vec{S}$, with gyromagnetic ratio $g_\mu = 2$, where $\vec{S}$ is the spin. However, quantum loop corrections yield small deviations from $g_\mu = 2$, which are parameterized by the anomalous magnetic moment \cite{Lindner:2016bgg}. A scalar particle shifts $\Delta a_\ell$ by,
\begin{equation}
\Delta a_{\ell}= 
\frac{(g - 2)_{\ell}}{2} \equiv \frac{\lambda^2}{8\pi^2} \int_0^1 dz \, \frac{(1 - z)^{2} (1+z)}{(1 - z)^{2} + z (\sfrac{m_S}{m_\ell})^{2}},
\end{equation}

where $\ell = e, \, \mu$, and $\lambda$ is the coupling strength between the new scalar particle with leptons.
The anomalous magnetic moment can be accurately measured taking advantage of the principle of Larmor precession, whose frequency is proportional to the magnetic field which the charged particle is immersed in \cite{Lindner:2016bgg, Muong-2:2023cdq}. In a similar vein, the electron anomalous magnetic moment has been constrained to be $\Delta a_e^{\mathrm{Rb}} \equiv a_e^{\mathrm{Exp}, \mathrm{Rb}}-a_e^{\mathrm{SM}}=(4.8 \pm 3.0) \times 10^{-13}$ \cite{Hanneke:2008tm}. The agreement between the SM prediction and the observations is much better. We will adopt the central value $\Delta a_\mu=a_\mu^{exp}-a_\mu^{SM}=39\times 10^{-11}$ reported in \cite{Li:2025myw}, and the central value of $\Delta a_{e}$ reported in \cite{Hanneke:2008tm} from hereafter, to set our constraints on the plane $\lambda$-$m_{S}$.

\section{Belle data analysis}
\label{belledata}
The BELLE collaboration reports results about the search for a new leptophilic scalar $S$ \cite{Belle:2022gbl}. The data used in this analysis was recorded by the BELLE experiment, which was operated until June 2010,  from the collision of 8 GeV electrons with 3.5 GeV positrons at the KEKB collider. The data-set corresponds to a luminosity of 626 $fb^{-1}$ collected after the upgrade of the silicon vertex detector in October 2003. The processes analyzed by the collaboration were ${e}^{+}{e}^{-}\rightarrow{\tau}^{+}{\tau}^{-}{\ell}^{+}{\ell}^{-}$, for which the number of events were measured for a given invariant mass of a pair of electrons or muons ($\ell^{\pm}$=$e^{\pm},\mu^{\pm}$). For these processes, a resonance of the invariant mass distribution compatible with the new scalar mass was expected; however, no signal was seen different from the background. In the absence of a signal, the BELLE collaboration imposed upper limits on the cross section of the asymmetric collision $\sigma(e^{+}e^{-} \rightarrow \tau^{+}\tau^{-}S$, $S \rightarrow e^{+}e^{-}(\mu^{+}\mu^{-}))$ at the 90$\%$ confidence level \cite{Belle:2022gbl}. We emphasize that this analysis was based on Eq.\eqref{effective}. We are interested in exploring a different setup, as described in the next section.

\section{Our analysis}
\label{ouranalysis}
Following this strategy and using the data analysis from the BELLE collaboration, we computed the production cross-section $\sigma(e^{+}e^{-} \rightarrow \tau^{+}\tau^{-}S$, $S \rightarrow e^{+}e^{-}(\mu^{+}\mu^{-}))$ for a subtly different model, where the leptophilic scalar $S$ couples uniquely and universally with the charged leptons of the standard model. The Lagrangian considered is given by,

\begin{equation}
    \mathcal{L}_{eff}=\frac{1}{2}(\partial_{\mu} S)^2-\frac{1}{2}m_{S}^{2}S^{2}+\sum_{\ell=e,\mu,\tau} \lambda \bar{\ell}  \ell S ,
    \label{lagrangian_leptophilic}
\end{equation}where $\lambda$ is a free parameter representing the universal coupling of the scalar $S$ with the SM-leptons. A coupling of this nature can arise if the new scalar does not develop a vacuum expectation value. In other words, it is an inert scalar. This lagrangian is often taken as a benchmark in experimental searches but one should bear in mind that is not invariant under the $SU(2)_L \times U(1)_Y$ gauge symmetry, but it could arise from effective operators of dimension-5 of the type $\frac{1}{\Lambda}\bar{L}\, H\, e_R\, S$, where $\Lambda$ would be the associated scale of new physics \cite{Chen:2018vkr}. Anyway, we will set aside any particular UV completion to focus on a more model-independent approach based on the simplified lagrangian Eq.\eqref{lagrangian_leptophilic}.
Moreover, we will assume that this inert scalar has only diagonal couplings to avoid dangerous flavour-changing interactions \cite{Lindner:2016bgg}. Hence, the inert scalar can decay into fermion pairs and photons with the following decay width,

\begin{align}
\Gamma_{S} =\sum_{i=e,\mu,\tau} \frac{\lambda^2 m_{S}}{8\pi}\left(1 - r_i\right)^{\,3/2}\;\nonumber\\
+ \frac{\alpha^{2} m_S^{3}}{256 \pi^{3}}
\left| \sum_{i = e,\mu,\tau} \frac{\lambda}{m_i} \, F_{1/2}(r_i,0) \right|^{2},
\label{Gamma} 
\end{align}
where $r_i = \sfrac{4 m_i^2}{m_S^2}$ and $F_{1/2}$ is defined in the Appendix.

With this information, we have implemented the model \eqref{lagrangian_leptophilic} in the Feynrules package, and then exported the UFO (Universal Feynman Output) files \cite{Alloul:2013bka}.  These UFO files were then used by the Madgraph Monte Carlo event generator to calculate the cross section of the processes of interest $\sigma(e^{+}e^{-} \rightarrow \tau^{+}\tau^{-}S$, $S \rightarrow e^{+}e^{-}(\mu^{+}\mu^{-}))$.

\begin{figure}
    \centering
    \includegraphics[width=0.99\linewidth]{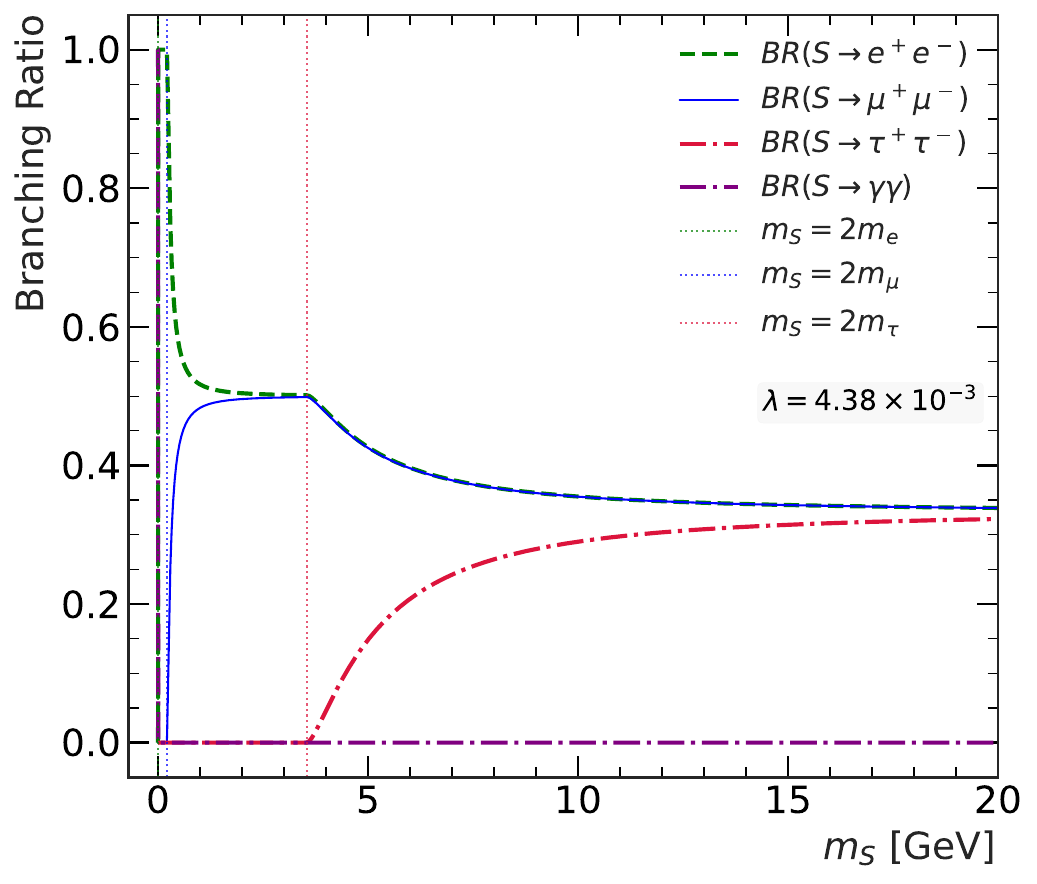}
    \includegraphics[width=0.99\linewidth]{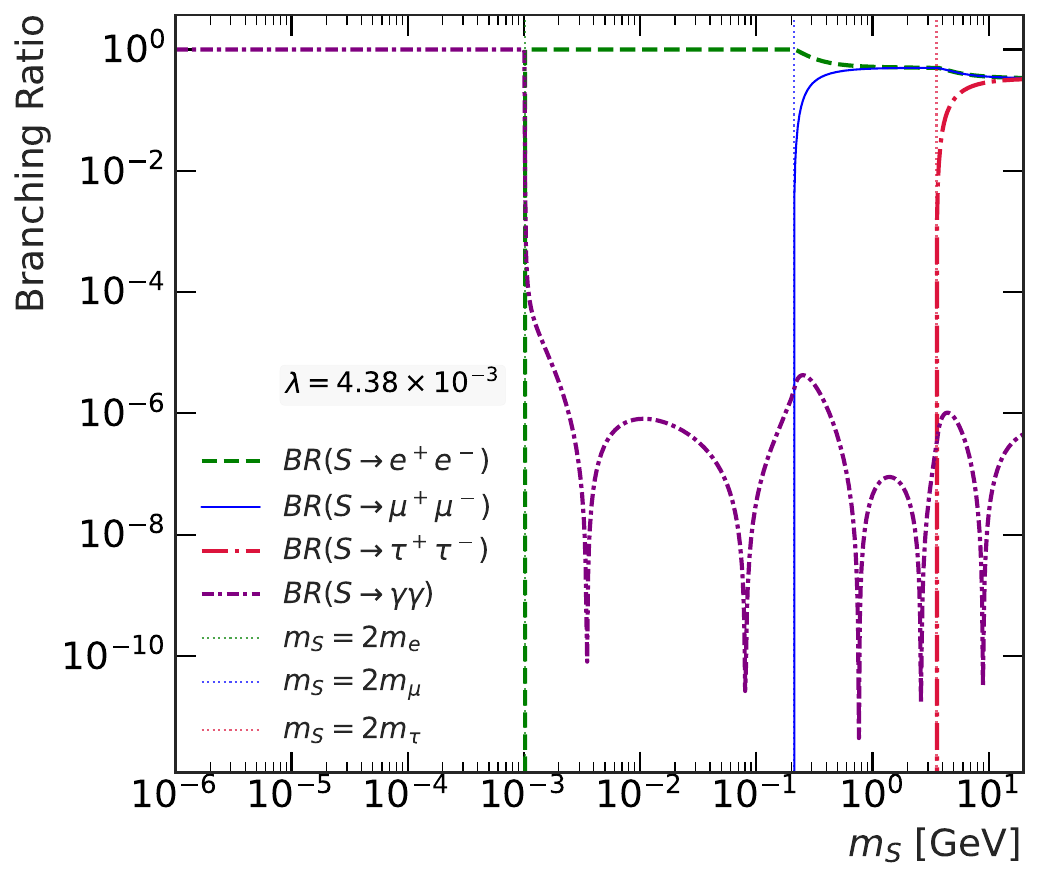}
    \caption{Branching ratios of the leptophilic scalar $S$ governed by the Lagrangian \eqref{lagrangian_leptophilic} for $\lambda=4.3 \times 10^{-3}$. The top panel is in linear scale, the bottom panel is in a log-log scale to capture in greater detail the kinematic transitions and hierarchies of the channels over a broader range ($m_{S} \sim 10^{-6} - 20 \, \text{GeV}$).\label{Brs}}
\end{figure}

Before explicitly explaining our analysis, we briefly discuss the properties of the branching ratios of this scalar $S$ derived from Eq. \eqref{Gamma}. The top panel in Fig.~\eqref{Brs} shows the branching ratios of the decays of $S$ as a function of its mass $m_{S}$ in linear scale for an arbitrary $\lambda$ value, and $m_{S}$ ranging from $10^{-4}$ to 20 GeV, covering all the region where the collider analysis has been performed ($m_{S} \sim 0.04 - 6.5 \, \text{GeV}$). Notice that the decay into  $\gamma\gamma$ is small reflecting the loop nature of the processes. The bottom panel of Fig.~\eqref{Brs} uses a log-log scale to capture in greater detail the kinematic transitions and hierarchies of the channels over a broader range ($m_{S} \sim 10^{-6} - 20 \, \text{GeV}$). In this bottom panel, it is clear that the $S \to \gamma\gamma$ channel dominates for $m_{S} \ll 2 m_e$ due to kinematics. The channel $S \to e^+e^-$ dominates for masses near $1 \, \text{GeV}$, while $S \to \mu^+\mu^-$ starts competing once $m_{S} > 2m_\mu$. For higher masses, $m_{S} > 2m_\tau$, the channel $S \to \tau^+\tau^-$ becomes sizable. It is important to highlight how these branchings differ from the mass-dependent coupling case, where the branching ratio to muons becomes nearly one above the $2m_{\mu}$ threshold, while for the universal couplings the branching ratios to muons and electrons are competitive. The oscillations observed in $S \to \gamma\gamma$ are due to interferences in the loop calculation associated with the terms in Eq.~\eqref{Gamma}.

\begin{figure}[ht]
    \begin{minipage}[b]{0.5\textwidth}
                \includegraphics[width=\textwidth]{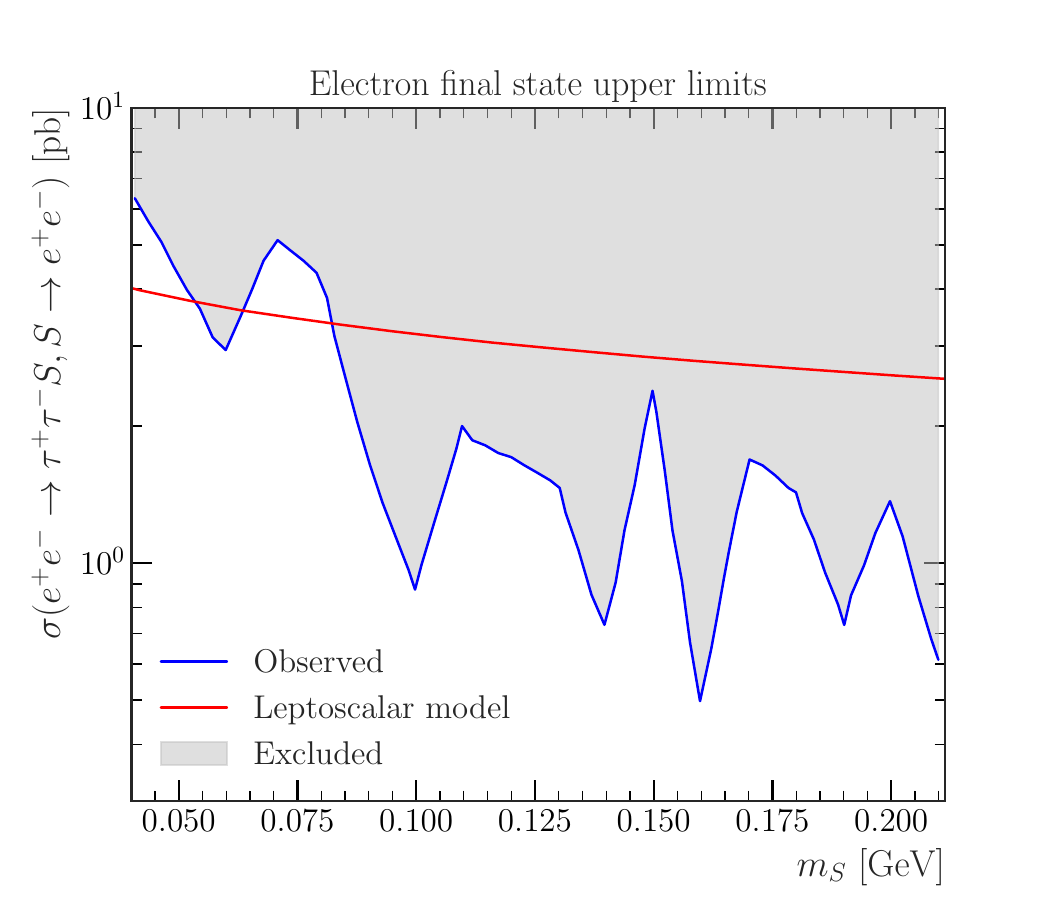}
        \caption{Production cross section of the light scalar, ${e}^{+}{e}^{-}\rightarrow{\tau}^{+}{\tau}^ {-}S$, with the scalar decaying into ${e}^{+}{e}^{-}$, for $\lambda =4.38 \times 10^{-3}$.}
        \label{excluded_electron}
    \end{minipage}
    \hfill
\end{figure}

Now, we will discuss in detail how we carried out our analysis in light of BELLE's results. We have computed the cross section $\sigma(e^{+}e^{-} \rightarrow \tau^{+}\tau^{-}S$, $S \rightarrow e^{+}e^{-}(\mu^{+}\mu^{-}))$ as a function of $m_{S}$ and $\lambda$, and compared with the upper limit reported in \cite{Belle:2022gbl}. For concreteness, we plotted this cross section in Fig.\eqref{excluded_electron} for $\lambda=4.38 \times 10^{-3}$. In Fig.\eqref{excluded_electron} the blue and red line correspond to the experimental (upper limit derived by BELLE through $e^{+}e^{-} \rightarrow \tau^{+}\tau^{-}S, S \rightarrow e^{+}e^{-}$ channel) and the simulated cross section driven by Eq.\eqref{lagrangian_leptophilic}, respectively. The gray region, the region above the blue line, is the region where no signal apart from the background was observed in the BELLE analysis, therefore, any simulated cross section value above the blue line corresponds to an excluded region by the experiment. Then, we conclude that for this channel and $\lambda$ value, $53MeV \leq m_{S} \leq 63MeV$ and $m_{S} \geq 83$~MeV are excluded. This exclusion limit is highly sensitive to the Yukawa coupling $\lambda$ because the smaller $\lambda$ the smaller the production cross section. To determine the upper limits through $e^{+}e^{-} \rightarrow \tau^{+}\tau^{-}S, S \rightarrow \mu^{+}\mu^{-}$ channel, a similar procedure is followed. However, it should be noted that the presence of meson resonances in this mass range, in particular the significant peaks observed in regions near the nominal masses of J/$\psi$ and $\psi$(2S), reduces the sensitivity of the BELLE experiment. For this reason, in the original BELLE analysis, a window of $\pm$50 MeV around these mass values was excluded. 
%

\begin{figure}[h!]
    \centering
        \includegraphics[width=1.0\linewidth]{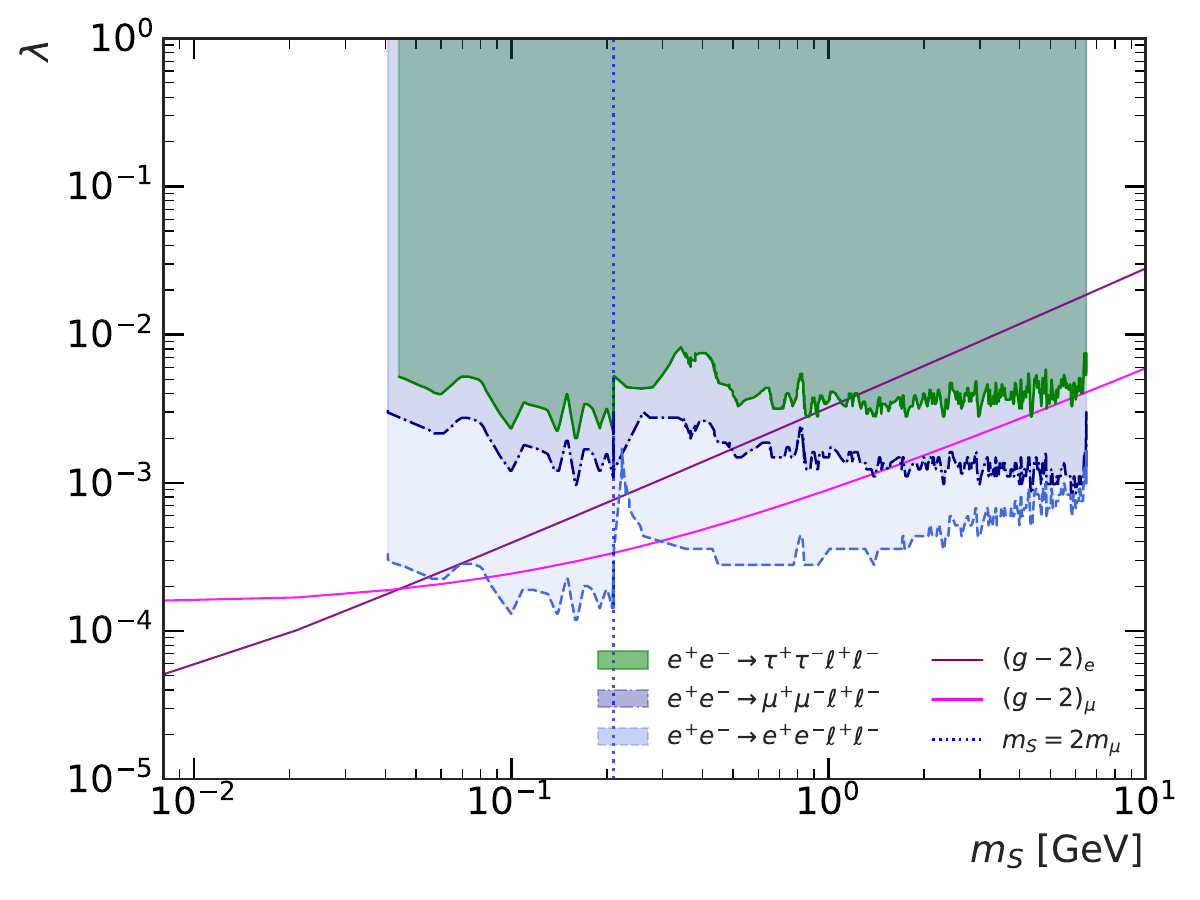}
    \caption{The shaded green region represents the exclusion region} from BELLE on the leptophilic scalar with universal couplings to leptons, ($\tau^{+}\tau^{-}S$) case. The blue regions represent the excluded regions derived using Eq.\eqref{rescale}, for the ($\mu^{+}\mu^{-}S$), ($e^{+}e^{-}S$) cases. The solid purple and pink lines represent the bounds stemming from the electron and muon anomalous magnetic moment. Assuming them to be consistent with SM predictions, points above these solid lines are excluded.
    \label{excluded_region}
\end{figure}

Following this reasoning, for different values of $\lambda$, we were able to compare the theoretical prediction of the production cross-section $\sigma(e^{+}e^{-} \rightarrow \tau^{+}\tau^{-}S, S \rightarrow e^{+}e^{-}(\mu^{+}\mu^{-}))$ with the data published by the Belle Collaboration, allowing us to separate excluded and allowed regions according to the data and place lower mass limits on the leptophilic scalar $S$. The exclusion regions obtained for this channel can be seen in Fig.\eqref{excluded_region} (green region). For this particular case ($\tau^{+}\tau^{-}S$), the data acquisition and event selection criteria performed by the BELLE collaboration consists of selecting only events that have four tracks associated with four charged leptons with zero net charge. Of these four tracks, two are required to have a common vertex (to reconstruct the hypothetical scalar $S$) and the other two are required to be tau leptons, originally motivated by the Lagrangian of Eq.\eqref{effective}. As mentioned in the introduction, the decay channel $S \rightarrow e^+e^-$ was considered for $ 2 m_e < m_{S} \leq 2m_\mu$, while the $S$ decay into muon pairs was considered for $ 2 m_\mu < m_{S} \leq 6.5$~GeV. Nevertheless, in our case the couplings of the scalar $S$ with leptons are universal, and we are not subject to the constraint that the coupling of the scalar $S$ with the $\tau$ leptons is the dominant one, making the processes $e^+e^- \rightarrow \mu^{+}\mu^{-}e^{+}e^{-}(\mu^{+}\mu^{-})$ and $e^+e^- \rightarrow e^{+}e^{-}e^{+}e^{-}(\mu^{+}\mu^{-})$ equally important to study. These processes should allow for more sensitive searches due to the availability of final-state electrons and muons (compared to $\tau$ leptons wich are difficult to reconstruct experimentally). As mentioned above, the BELLE detector is characterized by its excellent resolution for this type of channels. In order to implement the analysis for these new processes, we assumed that the number of events observed in these processes should be at least equal to the number of events observed in the original ($\tau^{+}\tau^{-}S$) selection criteria (this is because tau lepton efficiency is usually lower than the $e/\mu$ efficiency). Then, we can rescale the cross section measured by the BELLE collaboration, as shown in the following equation:

\begin{equation}
    \sigma(e^{+}e^{-} \rightarrow {\ell}_{i}^{+}{\ell}_{i}^{-}S) = \frac{{\epsilon}_{eff}^{\tau}}{{\epsilon}_{eff}^{{\ell}_{i}}}\times\sigma(e^{+}e^{-} \rightarrow {\tau}^{+}{\tau}^{-}S),
    \label{rescale}
\end{equation}

where $S$ continues to decay into ($e^{+}e^{-}$) pairs for $ 2 m_e < m_{S} \leq 2m_\mu$, and into  $(\mu^{+}\mu^{-})$ pairs for $ 2 m_\mu < m_{S} \leq 6.5$~GeV. Here ${\epsilon}_{eff}^{{\ell}_{i}}$ are the electron or muon efficiencies, whose values were taken as 0.86 and 0.88, respectively \cite{Belle-II:2022fsw}, and ${\epsilon}_{eff}^{{\tau}}$ is the tau lepton reconstruction efficiency, where we have assumed a global value of 0.7 to be conservative. The new exclusion regions are shown in Fig.\eqref{excluded_region} (blue regions). 


In FIG.\eqref{excluded_region}, we superimpose our limits delimited by shaded areas with those rising from lepton magnetic moments \cite{Muong-2:2025xyk, Li:2025myw, Hanneke:2008tm}. We remind the reader that we imposed $\Delta a_\mu$ and $\Delta a_e$ to reproduce the central value. As the discrepancies apparently died off, one can use them to constrain new physics, as exhibited in FIG.\eqref{excluded_region}.

The importance of our work is demonstrated by the blue regions in  FIG.\ref{excluded_region}. The original limit from BELLE (green region) represents the exclusion region on the leptophilic scalar with universal couplings to leptons, ($\tau^{+}\tau^{-}S$) case, originally inspired by Eq.\eqref{effective}. Notice that in this case BELLE yields constraints stronger than the electron g-2 only for $m_S > 1$~GeV. The muon g-2 limit is stronger than the BELLE's original bound across the entire parameter space. However, as we are investigating a sub-GeV scalar with universal couplings to the SM-leptons there is no need to focus on the ($\tau^{+}\tau^{-}S$) processes. Taking into account the others ($\mu^{+}\mu^{-}S$), ($e^{+}e^{-}S$) channels, assuming an equal number of events in the three channels and the efficiencies reported by BELLE collaboration, we show in  FIG.\ref{excluded_region} that BELLE can provide stronger constraints than measurements of the magnetic moments.\footnote{We did not include bounds based on the $\tau$ magnetic moment because they are too weak \cite{Lindner:2016bgg}} The $e^+e^- \rightarrow \mu^+\mu^- e^{+}e^{-}(\mu^{+}\mu^{-})$ processes yields a bound stronger than the muon magnetic moment only for $m_S> 2$~GeV. On the other hand, the $e^+e^- \rightarrow e^+e^- e^{+}e^{-}(\mu^{+}\mu^{-})$ channel is found to be more constraining than the ($\mu^{+}\mu^{-}S$) and ($\tau^{+}\tau^{-}S$) channels, and more stringent than those stemming from magnetic moment probes nearly across the entire mass range from $m_S= 100$~MeV to $6.5$GeV. This result highlights the importance of our analysis and the need to consider different new physics setups.

In summary, if BELLE were to perform a more general analysis considering all three production channels and detailed detector effects, the experiment could be an excellent laboratory for investigating new physics in the sub-GeV mass range, as suggested by our results in Fig.\eqref{excluded_region}. We emphasize that our findings are applicable to light scalars that couple to leptons with equal strength according to Eq.\eqref{lagrangian_leptophilic}.

\section{Conclusions}
\label{conclusions}
Light scalars coupling to leptons has been invoked in several new physics constructions. Collaborations typically assume lepton-specific couplings. In this case, the light scalar $S$ couples stronger with tau-leptons. Assuming a leptophilic scalar with universal coupling to leptons, the scalar that couples equally to all charged leptons, we computed the production cross section $\sigma(e^{+}e^{-} \rightarrow \tau^{+}\tau^{-}S, S \rightarrow e^{+}e^{-}(\mu^{+}\mu^{-}))$, and compared our results with the experimental data from Belle. Taking into account the detector efficiencies, we extended this analysis to incorporate the $e^+e^- \rightarrow \mu^+\mu^- e^{+}e^{-}(\mu^{+}\mu^{-})$ and $e^+e^- \rightarrow e^+e^- e^{+}e^{-}(\mu^{+}\mu^{-})$ production channels. We showed that the $e^+e^- \rightarrow e^+e^- e^{+}e^{-}(\mu^{+}\mu^{-})$ is more constraining and the BELLE sensitivity is greater than the one coming from electron-muon magnetic moment measurements assuming them to be consistent with SM predictions. In particular, for this more restrictive channel, we can solidly impose $\lambda < 1.2 \times 10^{-4}$ across the mass range 40~${\rm MeV} < m_S < 1$~GeV, weakening a bit toward larger masses. Therefore, if one only considers the ($\tau^{+}\tau^{-}S)$ channel, inspired by the flavor-dependent couplings Eq.\eqref{effective}, BELLE indeed yields weaker constraints than those arising from g-2, but in a flavor-independent scenario, the situation changes, and Belle stands as a promising laboratory for sub-GeV new physics as suggested by our findings in FIG.\eqref{excluded_region}.
\section*{Acknowledgements}

We thank Jacinto Neto and Carlos Pires for discussions. The work of Y.M. Oviedo-Torres and J. Zamora-Saa was funded by ANID – Millennium Science Initiative Program – ICN2019\_044. J. Zamora-Saa was partially supported by FONDECYT grant 1240216 and 1240066. YV is supported by FAPESP grants no. 2018/25225-9 and 2023/01197-4.Y.M. OviedoTorres was supported by FONDECYT grant 3250068. FSQ is supported by Simons Foundation (Award Number:1023171-RC), FAPESP Grant 2018/25225-9, 2021/01089-1, 2023/01197-4, ICTP-SAIFR FAPESP Grants 2021/14335-0, CNPq Grants 307130/2021-5 and 403521/2024-6, PROPESq-UFRN grant 758/2023, ANID-Millennium Science Initiative Program ICN2019\_044, PROPESq-UFRN grant 758/2023, and IIF-FINEP grant 213/2024. We thank IIP for the local cluster bulletcluster which was instrumental in this work.

\section{Appendix}

In this appendix, we present the loop function relevant for the S decay into $\gamma\gamma$. 
The function $F_{1/2}$ used in the inert scalar decay rate can be derived using the Package X program \cite{Patel:2015tea,Chen:2018vkr},  

\begin{equation}
F_{1/2}(r_i, 0) =
-2r_i \left[1 + (1-r_i) \arcsin^2\!\left(\frac{1}{\sqrt{r_i}}\right)\right]
\end{equation}for $r_i \geq 1$ or,
\begin{equation}
F_{1/2}(r_i, 0) = -2r_i \left[1 - \frac{1-r_i}{4}
\left(\log \frac{1+\sqrt{1-r_i}}{1-\sqrt{1-r_i}} - i\pi\right)^{2}\right]
\end{equation} for $r_i < 1$.

\section*{References}
\bibliographystyle{apsrev4-1}
\bibliography{references}
\end{document}

%% file: Diagrams.tex
\usepackage[scientific-notation=true]{siunitx}
\usepackage{float, verbatim, amsmath, amssymb, fullpage, paralist, cancel, listings, subfig, enumitem, siunitx, graphicx, xcolor, tikz, booktabs, tcolorbox, textcomp, url, parse lines, fontawesome5, marginnote, smartdiagram, forest, pifont, pgfcalendar, pgfgantt, fix-cm, ulem, soul, multirow, xfrac, colortbl, xspace, fontenc, helvet, mathptmx,ragged2e,csquotes} 
\definecolor{smcolor}{rgb}{0.8,0.5,0.5}
\definecolor{bsmcolor}{rgb}{0.4,0.6,0.8}
\definecolor{pvcolor}{rgb}{0.8,0.8,0.8}
\definecolor{cobalt}{rgb}{0.0, 0.28, 0.67}
\definecolor{darkelectricblue}{rgb}{0.33, 0.41, 0.47}
\definecolor{darkpowderblue}{rgb}{0.0, 0.2, 0.6}
\definecolor{darktangerine}{rgb}{1.0, 0.66, 0.07}
\definecolor{pastelviolet}{rgb}{0.8, 0.6, 0.79}
\definecolor{black}   {RGB}{0.0, 0.13, 0.28}
\definecolor{dukeblue}    {rgb}{0.0, 0.0, 0.61}
\definecolor{oxfordblue}{rgb}{0.0, 0.13, 0.28}
\definecolor{navyblue}{rgb}{0.0, 0.0, 0.5}
\definecolor{linkred}{RGB}{165,0,33}

\usesmartdiagramlibrary{additions}
\usepackage{hyperref}
\usepackage{cleveref}[capital]
\usepackage{textcase}
\hypersetup{
        linktocpage=true,
        setpagesize=true,
	colorlinks= true,
	linkcolor=cobalt!50!black,
        menucolor=cobalt!50!black,
	filecolor=cobalt!50!black,      
	urlcolor=cobalt!50!black,
        citecolor=cobalt!50!black,
        citebordercolor=cobalt!50!black,
	pdftitle={Light Dark Scalar},
        pdfsubject={Latex},
        pdfauthor={Yoxara Villamizar},
        pdfkeywords={BSM Phenomenology, SM, Collider, DM}
}
\graphicspath{{Figures/}} 
\newlist{firstitemize}{itemize}{1}
\setlist[firstitemize,1]{label=\color{cobalt!70!black} \ding{111}, nosep, leftmargin=*, before=\normalcolor}
\newlist{seconditemize}{itemize}{1}
\setlist[seconditemize,1]{label=\color{cobalt!70!black}\ding{226}\normalcolor, nosep, leftmargin=*} 
\newlist{thirditemize}{itemize}{1}
\setlist[thirditemize,1]{label=\color{cobalt!70!black} \ding{118}, nosep, leftmargin=*, before=\normalcolor}

\newcommand{\z}{\mathrm{Z}}

\renewcommand{\bar}{\overline}

\usepackage{tikz}
\usetikzlibrary{arrows.meta,decorations,calc,shadows.blur,decorations.pathreplacing,decorations.pathmorphing,decorations.markings,shapes, arrows, positioning,bending,chains,matrix, patterns,mindmap,trees}
\tikzset{
>=triangle 45,
scalar/.style={decorate, dashed, draw=cobalt!50!black, thick, line width=0.8pt}, 
photon/.style={decorate, line width=0.85pt, draw=black, decoration={snake, segment length=5, amplitude=3pt, post length=0.5mm, pre length=0.5mm}}, 
particle/.style={draw=black, postaction={decorate}},
fermion/.style={draw=black, line width=0.6pt, postaction={decorate}, decoration={markings,mark=at position .5 with {\arrow[draw=black,scale=1.2,>=stealth]{>}}}},
antifermion/.style={draw=black, line width=0.6pt, postaction={decorate},decoration={markings,mark=at position .5 with {\arrow[draw=black,scale=1.2,>=stealth]{<}}}},
}

\newcommand\eetottP{
\resizebox{\linewidth}{!}{
\begin{tikzpicture}
    \draw[fermion] (0,2.5) -- (1,1.5) node[at start, left, xshift=0.1cm] {\large $e^{-}$};
    \draw[antifermion] (0,0.5) -- (1,1.5) node[at start, left, xshift=0.1cm] {\large $e^{+}$};
    \draw[photon] (1,1.5) -- (2.2,1.5) node[midway, above, yshift=-0.00cm] {\large $\gamma/\z$};
    \draw[fermion] (2.2,1.5) -- (3.2,2.5) node[at end, right, xshift=-0.1cm] {\large $\tau^{-}$};
    \draw[antifermion] (2.2,1.5) -- (3.2,0.5) node[at end, right, xshift=-0.1cm] { \large $\tau^{+}$};
    \draw[scalar] (2.7,1) -- (3.8,1) node[midway, above, yshift=-0.0cm, cobalt!50!black] { \large $S$};
    \draw[fermion] (3.8,1) -- (4.5,1.5) node[at end, right, xshift=-0.1cm] {\large  $\ell^{-}$};
    \draw[antifermion] (3.8,1) -- (4.5,0.5) node[at end, right, xshift=-0.1cm] { \large $\ell^{+}$};
\end{tikzpicture}
}
}